 \documentstyle[prl,aps]{revtex}  
  \def\ack{\,|\,}

\begin{document}
\draft


 \twocolumn[\hsize\textwidth\columnwidth\hsize  
 \csname @twocolumnfalse\endcsname              

\title{
On the Backbending Mechanism of $^{48}$Cr
}
\author{Kenji Hara$^{(1)}$, Yang Sun$^{(2)}$ and Takahiro Mizusaki$^{(3)}$
}
\address{
$^{(1)}$Physik-Department, Technische Universit\"at M\"unchen,
D-85747 Garching, Germany\\
$^{(2)}$Department of Physics and Astronomy, University of Tennessee, 
Knoxville, TN 37996, USA\\
$^{(3)}$Department of Physics, University of Tokyo, Hongo, Tokyo 113,
Japan\\
}

\maketitle

\begin{abstract}
The mechanism of backbending in $^{48}$Cr is investigated in terms of
the Projected Shell Model and the Generator Coordinate Method. It is
shown that both methods are reasonable shell model truncation schemes.
These two quite different quantum mechanical approaches lead to a
similar conclusion that the backbending is due to a band crossing
involving an excited band which is built on simultaneously broken
neutron and proton pairs in the ``intruder'' subshell $f_{7/2}$. It is
pointed out that this type of band crossing is usually known to cause
the second backbending in rare-earth nuclei.
\end{abstract}



 ]  


Investigation of the yrast band in the $^{48}$Cr nucleus has recently
become a particularly interesting subject in nuclear structure studies
because of the full $pf$-shell model calculation \cite{fpshell} on the
one hand and of the spectroscopic measurements \cite{exp_a,exp_b}
on the other. It is a light nucleus for which an exact shell model
diagonalization is feasible, yet exhibits remarkable high-spin phenomena
usually observed in heavy nuclei: large deformation, typical rotational
spectrum and the backbending in which regular rotational spectrum is
disturbed by a sudden irregularity at a certain spin. This nucleus is
therefore an excellent example for theoretical studies, providing a
unique testing ground for various approaches.
 
The intrinsic and laboratory frame descriptions of this nucleus were
presented in a paper by the Strasbourg-Madrid collaboration
\cite{fpshell}. These approaches are complementary views of the same
problem from two extremes. On the one hand, the cranked
Hartree-Fock-Bogoliubov (CHFB) description interprets the problem in
terms of the intrinsic frame on which the lowest rotational band (yrast
band) is built. It can provide a nice physical insight but does not
treat the angular momentum as a good quantum number. On other hand, the
$pf$-shell model (pf-SM) approach solves the problem fully quantum
mechanically and provides the exact solution of the Hamiltonian within
the $pf$-shell. However, in the pf-SM, a single shell model
configuration does not correspond to any excitation mode of deformed
nucleus and therefore millions of many-body basis states are necessary
even to represent the lowest eigenstate of the Hamiltonian.
Consequently, the physical insight is lost and interpretation of the
result becomes very difficult. The purpose of the present work is to
clarify the physics associated with the yrast spectrum of the nucleus
$^{48}$Cr from two different quantum mechanical view points.

To extract physics, it is desirable to use a shell model basis which has
a good classification scheme in the sense that a simple configuration
corresponds (approximately) to a low excitation mode of the nucleus.
This suggests us to use a deformed basis corresponding to the optimal
set of basis states. In fact, a basis truncation can be most easily done
by selecting low-lying states if a proper deformed basis is used. To
carry out a shell model type calculation with such a basis, the broken
rotational symmetry (and the particle number conservation if necessary)
has to be restored. This can be done by using the projection method to
form a many-body basis in the laboratory frame. After this procedure,
one diagonalizes the Hamiltonian. Such an approach lies conceptually
between the two extreme methods mentioned above and takes the advantages
of both. This is exactly the philosophy on which the Projected Shell
Model (PSM) \cite{review} is based.

The PSM uses the Nilsson+BCS representation as the deformed
quasiparticle (qp) basis. Before performing a calculation for $^{48}$Cr,
one has to find out where the optimal basis is. The experimental
lifetime measurement \cite{exp_b} suggests an axially symmetric
deformation of $\beta \approx 0.28$ near the ground state of $^{48}$Cr,
which roughly corresponds to $\varepsilon_2 = 0.25$. We simply take this
information and build our shell model basis at this deformation.

The set of multi-qp states relevant for our shell model configuration
space is
\begin{eqnarray}
\ack\Phi_\kappa\rangle=\{ \ack 0 \rangle,
\ a^\dagger_{\nu_1}a^\dagger_{\nu_2}\ack 0\rangle,
\ a^\dagger_{\pi_1}a^\dagger_{\pi_2}\ack 0\rangle,
\ a^\dagger_{\nu_1}a^\dagger_{\nu_2}a^\dagger_{\pi_1}
a^\dagger_{\pi_2}\ack 0\rangle \},
\label{conf}
\end{eqnarray}
where $a^\dagger$'s are the qp creation operators, $\nu$'s ($\pi$'s)
denote the neutron (proton) Nilsson quantum numbers which run over
properly selected (low-lying) orbitals and $\ack 0 \rangle$ the
Nilsson+BCS vacuum or 0-qp state.

As in the usual PSM calculations, we will use the Hamiltonian
\cite{review}
\begin{equation}
\hat H = \hat H_0 - {1 \over 2} \chi \sum_\mu \hat Q^\dagger_\mu
\hat Q^{}_\mu - G_M \hat P^\dagger \hat P - G_Q \sum_\mu \hat
P^\dagger_\mu\hat P^{}_\mu,
\label{hamham}
\end{equation}
where $\hat H_0$ is the spherical single-particle Hamiltonian which
in particular contains a proper spin-orbit force, whose strengths (i.e.
the Nilsson parameters $\kappa$ and $\mu$) are taken from Ref.
\cite{brkm}. 
The second term in the Hamiltonian is the Q-Q interaction and the last
two terms the monopole and quadrupole pairing interactions,
respectively. It was shown \cite{Zuker96} that these interactions
simulate the essence of the most important correlations in nuclei, so
that even the realistic force has to contain at least these components
implicitly in order for it to work successfully in the structure
calculations. The interaction strengths are determined as follows: the
Q-Q interaction strength $\chi$ is adjusted by the self-consistent
relation such that the input quadrupole deformation $\varepsilon_2$ and
the one resulting from the HFB procedure coincide with each other
\cite{review}. The monopole pairing strength $G_M$ is taken to be
$G_M=\left[22.5-18.0(N-Z)/A\right]/A$ for neutrons and $G_M=22.5/A$ for
protons, which was first introduced in Ref. \cite{pairing}. This choice
of $G_M$ seems to be appropriate for the single-particle space employed
in the present calculation in which three major shells ($N=1,2,3$) are
used for both neutron and proton. Finally, the quadrupole pairing
strength $G_Q$ is assumed to be proportional to $G_M$, the
proportionality constant, which is usually taken in the range of 0.16 --
0.20, being fixed to 0.20 in the present work.

The eigenvalue equation of the PSM for a given spin $I$ takes the
form \cite{review}
\begin{equation}
\sum_{\kappa'}\left\{H^I_{\kappa\kappa'}-E^IN^I_{\kappa\kappa'}\right\}
F^I_{\kappa'}=0, 
\label{psmeq}
\end{equation}
where the Hamiltonian and norm matrix elements are respectively defined
by
\begin{equation}
H^I_{\kappa\kappa'}=\langle\Phi_\kappa\ack\hat H\hat P^I_{KK'}\ack
\Phi_{\kappa'}\rangle,~~N^I_{\kappa\kappa'}=\langle\Phi_\kappa\ack\hat
P^I_{KK'}\ack\Phi_{\kappa'}\rangle,
\label{elem}
\end{equation}
and $\hat P^I_{MK}$ is the angular momentum projection operator.
The expectation value of the Hamiltonian with respect to a ``rotational
band $\kappa$'' $H^I_{\kappa\kappa}/N^I_{\kappa\kappa}$ is called
a band energy. When they are plotted as functions of spin $I$, we call
it a band diagram \cite{review}. It will provide us a useful tool for
interpreting the result.

We have carried out not only the PSM calculation but also a calculation
based on the Generator Coordinate Method (GCM) \cite{mizu} with the same
Hamiltonian as used in the pf-SM \cite{fpshell}. This is because the PSM
Hamiltonian is quite schematic as it stands, so that we felt it
necessary to confirm the result by another theory using the same
Hamiltonian as the pf-SM. The relation between the PSM and GCM will be
discussed later. We will first explain briefly how the GCM is performed.

For given quadrupole moment $q_\mu$ and spin $I$, we look for the
minimum of
\begin{equation}
\langle \hat H' \rangle \equiv \langle \hat H \rangle 
+ c_1 \sum_{\mu=0,\pm2} (\langle \hat Q_\mu \rangle - q_\mu)^2
+ c_2 [\langle \hat J_x \rangle - \sqrt{I(I+1)} ]^2, 
\label{gcm}
\end{equation}
where $c_1$ and $c_2$ are predefined positive constants. This procedure
generates a constrained Hartree-Fock (CHF) state $\ack q,\gamma,I \rangle$, 
where
\begin{equation}
q_0=\sqrt{{5 \over {4\pi}}}q\cos\gamma,~~~q_{\pm2}=\sqrt{{5 \over
{8\pi}}}q\sin\gamma. 
\label{q-gamma}
\end{equation}
It is useful to plot an energy surface $\langle q,\gamma,I \ack \hat H
\ack q,\gamma,I \rangle$ in the $q$-$\gamma$ plane for each $I$, which
usually shows several local minima. It will help us interpreting the
result as we will soon see.
We then project such Slater determinants corresponding to various $q$
and $\gamma$ onto a good angular momentum $I$ and diagonalize the
Hamiltonian with them, the eigenvalue equation being again of the form
Eq.(\ref{psmeq}) with $\ack\Phi_\kappa\rangle=\{\ack q,\gamma,I\rangle\}$.
The total number of mesh points in the $q$-$\gamma$ parameter space is
66 in the present calculation, so that the size of the eigenvalue
equation for a given spin $I$ is at most $66*(2I+1)$ (some of them will
be discarded due to vanishing norm). This is another way of truncating
the shell model basis \cite{mizu}.

In Fig. 1, the results of the PSM and GCM for the $\gamma$-ray energy
$E_\gamma=E(I)-E(I-2)$ along the yrast band, together with that of
the pf-SM reported in Ref. \cite{fpshell}, are compared with the newest
experimental data \cite{exp_b}. One sees that four curves are
bunched together over the entire spin region, indicating an excellent
agreement of three theories with each other and with the data. The
sudden drop in $E_\gamma$ occurring around spin 10 and 12 corresponds to
the backbending in the yrast band of $^{48}$Cr.

In Fig. 2, three theoretical results for B(E2) are compared with the
data \cite{exp_b}. All the three theories use the same effective charges
(0.5e for neutrons and 1.5e for protons). 
Again, one sees that theories agree not only with
each other but also with the data quite well. The B(E2) values decrease
monotonously after spin 6 (where the first band crossing takes place in
the PSM, see Fig. 3 and discussions below). This implies a monotonous
decrease of the intrinsic Q-moment as a function of spin, reaching
finally the spherical regime at higher spins. This feature was
explicitly discussed in Ref. \cite{fpshell} within the CHFB framework.

Figs. 1 and 2 indicate that both PSM and GCM are reasonable shell model
truncation schemes as they reproduce the result of the pf-SM very well.
Let us study their band diagrams to understand why and how the
backbending occurs. As mentioned before, a band diagram displays band
energies of various configurations before they are mixed by the
diagonalization procedure Eq.(\ref{psmeq}). It can provide a transparent
picture of band crossings. Irregularity in a spectrum may appear if a
band is crossed by another one at certain spin. We remark in this
connection that a small crossing angle implies a smooth change in the
yrast band while a large crossing angle a sudden change, leading to a
backbending \cite{HS91}.

In Fig. 3, the band diagram of the PSM is shown. Different
configurations are distinguished by different types of lines, and the
filled circles represent the yrast states obtained after the
configuration mixing. Among several 2-qp bands which start at energies
of 2 -- 3 MeV, two of them (one solid and another dashed line) cross
the ground band at spin 6. They are neutron 2-qp and proton 2-qp bands
consisting of two $f_{7/2}$ quasiparticles of $\Omega=3/2$ and $5/2$
coupled to total $K=5/2-3/2=1$ forming the so-called $s$-band. The
crossing angle is relatively small so that the yrast band smoothly
changes its structure from the 0-qp to the 2-qp states around spin 6.
Therefore, no clear effect of this (first) band crossing is seen in the
yrast band (cf. Fig. 1). These $\Omega=3/2$ and $5/2$ Nilsson states are
nearly spherical in which the $j=7/2$ component dominates (95\% and 98\%
respectively). This is nothing other than the property which
characterizes the intruder states.

The above two ($K=1$) 2-qp bands can combine to a ($K=2$) 4-qp band
which represents simultaneously broken neutron and proton pairs. In Fig.
3, this 4-qp band (one of the dashed-dotted lines which becomes the
lowest band for $I \ge 10$) shows a unique behavior as a function of
spin. As spin increases, it goes down first but turns up at spin 6. This
behavior has its origin in the spin alignment of a decoupled band as
intensively discussed in Ref. \cite{review}. Because of this, it can
sharply cross the 2-qp bands between spin 8 and 10 and becomes the
lowest band thereafter, so that the yrast band gets the main component
from this 4-qp band. This is seen in the band diagram as a (second) band
crossing. Thus, we can interpret the backbending in $^{48}$Cr as a
consequence of the simultaneous breaking of the $f_{7/2}$ neutron and
proton pairs.

Similar band crossing picture emerges also from the band diagram of the
GCM (here we use the word ``band'' symbolically). In Fig. 4, two most
prominent bands in the GCM are shown together with the yrast band
obtained after the diagonalization. The one labelled as deformed band is
associated with a prolate minimum while the one labeled as spherical
band with the zero deformation which becomes a minimum only when spin is
higher. Fig. 4 shows the competition between these two bands which are
built by diagonalizing the Hamiltonian within small regions of the
$q$-$\gamma$ plane around the respective local minima in the energy
surface. Therefore, in the GCM, the backbending in $^{48}$Cr can be
interpreted as due to the crossing between the deformed and spherical
band. The latter dominates beyond spin 10 and this explains the sudden
decrease of the Q-moment in the pf-SM calculation.

While the deformed band in the GCM corresponds obviously to an admixture
of the 0-qp and a 2-qp band in the PSM, we note that the feature of the
spherical band in Fig. 4 looks similar to that of the 4-qp band in Fig.
3. In fact, the 4-qp band of the PSM can be considered as a spherical
band because the main part of its wavefunction is a product of the
$j=7/2$ components as mentioned before, while the spherical band of the
GCM can be thought as the 4-qp band because it consists mainly of the
$f_{7/2}$ neutrons and protons as confirmed by evaluating the occupation
number of each shell. We have thus the same physics in two different
languages and may conclude that the backbending of $^{48}$Cr is due to
the band crossing caused by a simultaneous neutron and proton pair
breaking in the $f_{7/2}$ shell. We remark that this statement differs
from that of a recent paper \cite{sakata} based on the CHFB which claims
that the backbending in $^{48}$Cr is not due to level crossing.

To conclude, both PSM and GCM are reasonable shell model truncation
schemes and suggest consistently that the backbending of $^{48}$Cr is
due to a band crossing. The PSM carries out the configuration mixing in
terms of qp-excitations while the GCM in terms of Slater determinants
that belong to different nuclear shapes. We have shown in the present
context that they describe the same physics. However, in the PSM
terminology, the backbending in $^{48}$Cr is not due to the crossing
between the 0-qp and 2-qp band (or the so-called $g$-$s$ band crossing)
but rather to the one between a 2-qp and 4-qp band. This is remarkable
because such a band crossing is known to lead to the second backbending
in rare-earth nuclei \cite{HS91}. The $g$-$s$ band crossing which
usually leads to the first backbending shows no prominent effect in
$^{48}$Cr except for the decrease in B(E2) values. This is because the
spin alignment of a 2-qp band cannot become large enough in light nuclei
as the maximal value is limited to $J=6$ in the intruder subshell
$f_{7/2}$, which is only the half of what is possible ($J=12$) in the
intruder subshell $i_{13/2}$ of rare-earth nuclei.

One basic question still remains. The PSM uses a schematic Hamiltonian
while the GCM the same realistic Hamiltonian as the pf-SM. How can the
PSM deliver a similar result as the latter two? The answer to this
question was actually given in Appendix B of Ref. \cite{review}. It was
proved that the band energies do not depend on details of the
Hamiltonian for the ground and intruder bands and that any (rotation
invariant) Hamiltonian, which gives similar values for (1) the
fluctuation of the angular momentum, (2) the Peierls-Yoccoz moment of
inertia and (3) the qp excitation energies, will lead essentially to the
same result. The first two conditions are the ground state properties
that determine the scaling of a band diagram while the last one the
relative position of various bands reflecting the shell filling of the
nucleus in question. The detailed spin dependence such as the signature
rules (for example, even/odd spins are favoured/unfavoured in an
even-even system) originate solely from the kinematics of angular
momenta. Note that the yrast band is essentially the envelope of the
band energies, cf. Fig. 3. This explains the reason why the PSM works so
nicely even with a simple schematic Hamiltonian.

Finally, we briefly comment on the GCM. Its wavefunction is a
superposition of various projected CHF states but, for heavier systems,
the constained HFB states should be used to take into account strong pairing
correlations more efficiently. Intuitively speaking, the CHF sates
corresponding to various nuclear shapes represent different moments of
inertia and they compete with one another, which is probably a new way
of interpreting the backbending. This formalism has an obvious advantage
that the normally deformed and superdeformed bands in one nucelus 
may be simultaneously described
on the same footing.

\medskip
The present work is supported in part by Grant-in-Aid for Scientific
Research (A)(2)(10304019) from the Ministry of Education, Science and
Culture of Japan.
K.H. and Y.S. acknowledge A.P. Zuker for a conversation made during the
{\em Drexel Shell Model Workshop} in Philadelphia, 1996, stimulating
them to carry out a PSM analysis \cite{unpublished} of the nucleus
$^{48}$Cr on which a part of the present work is based.

\baselineskip = 14pt
\bibliographystyle{unsrt}

\begin{thebibliography} {99}

\bibitem{fpshell}
E. Caurier, {\em et al}, Phys. Rev. Lett. {\bf 75}, 2466 (1995)
\bibitem{exp_a}
J.A. Cameron, {\em et al}, Phys. Lett. {\bf B387}, 266 (1996)
\bibitem{exp_b}
F. Brandolini, {\em et al}, Nucl. Phys. {\bf A642}, 387 (1998)
\bibitem{review}
K. Hara and Y. Sun, Int. J. Mod. Phys. {\bf E4}, 637 (1995)
\bibitem{brkm}
T. Bengtsson and I. Ragnarsson, Nucl. Phys. {\bf A436}, 14 (1985)
\bibitem{Zuker96}
M. Dufour and A.P. Zuker, Phys. Rev. {\bf C54}, 1641 (1996)
\bibitem{pairing}
W. Dieterich, {\em et al}, Nucl. Phys. {\bf A253}, 429 (1975)
\bibitem{mizu}
T. Mizusaki, {\em et al}, Phys. Rev. {\bf C59}, 1846R (1999)
\bibitem{HS91}
K. Hara and Y. Sun, Nucl. Phys. {\bf A529}, 445 (1991)
\bibitem{sakata}
T. Tanaka, K. Iwasawa and F. Sakata, Phys. Rev. {\bf C58}, 2765 (1998)
\bibitem{unpublished} K. Hara and Y. Sun, unpublished (1996)

\end{thebibliography}

\begin{figure}
\caption{ 
The $\gamma$-ray transition energies $E_\gamma=E(I)-E(I-2)$
as functions of spin. The experimental data is taken from Ref.
\protect\cite{exp_b} and the result of pf-SM from Ref.
\protect\cite{fpshell}. PSM and GCM are the present results.
}
\label{figure.1}
\end{figure}

\begin{figure}
\caption{ 
The B(E2) values as functions of spin. The experimental data are taken
from Ref. \protect\cite{exp_b} and the result of pf-SM from Ref.
\protect\cite{fpshell}. PSM and GCM are the present results.
}
\label{figure.2}
\end{figure}

\begin{figure}
\caption{ 
Band diagram of the PSM
}
\label{figure.3}
\end{figure}

\begin{figure}
\caption{ 
Band diagram of the GCM
}
\label{figure.4}
\end{figure}

\end{document}